\documentclass{jjap2}
\title{Single-crystal growth of the ternary BaFe$_2$As$_2$ phase using the vertical Bridgman technique}

\author{Rei \textsc{Morinaga}$^1$, Kittiwit \textsc{Matan}$^1$, Hiroyuki S. \textsc{Suzuki}$^2$ and Taku J. \textsc{Sato}$^1$\thanks{E-mail address: taku@issp.u-tokyo.ac.jp; corresponding author}}

\inst{$^1$Neutron Science Laboratory, Institute for Solid State Physics, University of Tokyo, 106-1 Shirakata, Tokai, Ibaraki 319-1106, Japan\\
$^2$ National Institute for Materials Science, 1-2-1 Sengen, Tsukuba, Ibaraki 305-0047, Japan}

\abst{Ternary Ba-Fe-As system has been studied to determine a primary solidification field of the BaFe$_2$As$_2$ phase.
We found that the BaFe$_2$As$_2$ phase most likely melts congruently and primarily solidifies either in the FeAs excess or Ba$_{x}$As$_{100-x}$ excess liquid.
Knowing the primary solidification field, we have performed the vertical Bridgman growth using the starting liquid composition of Ba$_{15}$Fe$_{42.5}$As$_{42.5}$.
Large single crystals of the typical size 10x4x2~mm$^3$ were obtained and their quality was confirmed by X-ray Laue and neutron diffraction.}

\kword{Fe-based superconductor, Ba-Fe-As system, Single-crystal growth}

\begin{document}
\maketitle

\section{Introduction} 
Recent discovery of the Fe-based superconductors attracts tremendous interests in the solid state physics community because of their wide elemental and structural variations, intriguing interplay of antiferromagnetic and structural ordering, and high superconducting transition temperatures up to about 52~K~\cite{kamihara06,kamihara08,ren08}.
The (Ba$_{1-x}$K$_x$)Fe$_2$As$_2$ compound is one of such Fe-based superconductors exhibiting the superconductivity at relatively high temperatures as $T_{\rm c} = 38$~K at $x = 0.4$~\cite{rotter08a,rotter08b}.

To reveal intrinsic superconducting properties of the (Ba$_{1-x}$K$_x$)Fe$_2$As$_2$ superconductor, high-quality and large-sized single crystals of both the superconducting K-doped and parent (non-doped) compounds are definitely needed.
Hence, shortly after the first report of the superconductivity, several single-crystal growths were reported~\cite{chen08,ni08,wang08,ronning08}.
Two strategies were found in the earlier reports; one is to use other elements, such as Sn, as flux, whereas the other is to use so-called self-flux method.
For the Sn flux growth, an extraordinary large amount of Sn is necessary to enable the single-crystal growth~\cite{ni08,ronning08}.
Furthermore, contamination of the resulting crystal by the flux element is always a serious issue.
On the other hand, in the self-flux growth~\cite{wang08}, as the initial liquid composition is rather closer to the target compound compared to the Sn flux growth, one may expect larger crystals.
Nevertheless, since the earlier study employs the slow cooling method, which generally has a size limitation of resulting crystals due to uncontrolled spontaneous nucleations, sizes of the obtained crystals were rather small for elaborated physical property research such as neutron scattering.
In addition, the melt composition ({\it i.e.}, the amount of flux) seems to be determined empirically; information on equilibrium phase diagram near the melting temperature, in particular the primary solidification field of the target compound, is highly desired to optimize the single-crystal growth.

In the present study, we have performed preliminary phase diagram investigation near the melting temperature to establish the primary solidification field of the BaFe$_2$As$_2$ phase from the Ba-Fe-As melt.
Based on the results, we have performed single-crystal growth of the ternary BaFe$_2$As$_2$ phase.
To overcome the size limitation of the previously applied slow-cooling method, the vertical Bridgman method was employed, where nucleation and growth proceed in a controlled manner.

\section{Experimental}
For the phase diagram determination, ternary alloys with several compositions near BaFe$_2$As$_2$ (= Ba$_{20}$Fe$_{40}$As$_{40}$) were prepared in the following manner.
Stoichiometric mixture of the elemental Ba (3N, chunk), Fe (-300mesh powder, 4N), and As (-200mesh powder, 5N), were put in an Al$_2$O$_3$ crucible, and sealed in a quartz tube under pure Ar atmosphere.
All the sample handling was carried out in a glove box filled with purified Ar gas.
Heat-treatments were performed using standard electric furnaces.
Typical heat-treatment procedure is as follows: the mixture was firstly heat-treated at 673~K for 6 hours, secondly at 873~K for 12~hours, and then the temperature was slowly increased to 1423~K or 1323~K at a rate 53~K/h.
The mixture was kept at the highest temperature at least for 0.5 hours, and then cooled down to the room temperature in the furnace.

The quenching experiment of the BaFe$_2$As$_2$ melt was performed using the induction furnace.
The pre-synthesized BaFe$_2$As$_2$ alloy was inserted into a high-purity Al$_2$O$_3$ crucible, and then sealed in a quartz tube under pure Ar atmosphere.
For this melting experiment, to protect the quartz tube, a silica wool was placed between the crucible and quartz tube as a thermal insulation.
The melt was confirmed by the eyes through a view port placed at the top of the quartz tube, as well as by observing melting heat by an infra-red thermometer.

Microstructures of the obtained alloys were checked by the scanning electron microscopy (SEM: JEOL JSM-5600), whereas compositions of the phases were determined by the energy dispersive X-ray analysis (EDX).
Crystal structures of the polycrystalline alloys were checked by the powder X-ray diffraction (Rigaku Miniflex).

Single crystals of BaFe$_2$As$_2$ were grown using the vertical Bridgman technique.
The Bridgman furnace used in the present study has a constant-temperature zone of longer than 100~mm, whereas the maximum temperature gradient of the growth zone is approximately 30~K/cm.
The polycrystalline Ba$_{15}$Fe$_{42.5}$As$_{42.5}$ alloy, prepared in the above manner, was loaded in a high-purity Al$_2$O$_3$ crucible with a bottom tapering to a point with an angle of 90$^{\circ}$ to obtain seed selection.
(The detail of the crucible may be found in the earlier report~\cite{sat98b}.)
The crucible was initially set in the constant-temperature zone, and heated up to 1443~K at a rate of 30~K/h.
The crucible was then gradually lowered at a speed of 0.5~mm/h.
Quality of the obtained crystal was checked by taking X-ray Laue backscattering diffractograms using a tungsten target, as well as the neutron diffraction taken at the ISSP-GPTAS triple-axis spectrometer installed at JRR-3, Tokai, Japan.
In addition, electrical resistivity of the obtained single crystal was measured using the conventional DC four-probe technique with an excitation current of 1~mA.

\section{Congruent or incongruent melting?}

\begin{figure}
\includegraphics[scale=0.28,angle=-90]{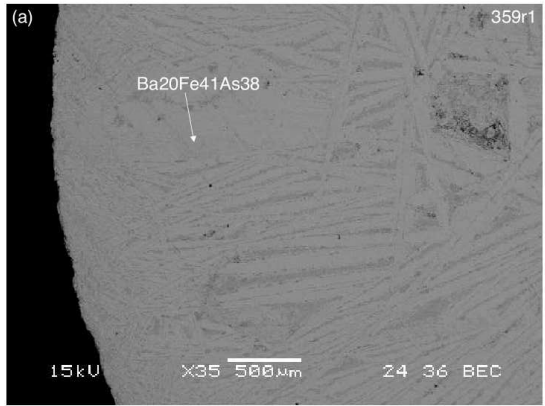}
\caption{Backscattered electron micrograph of the Ba$_{20}$Fe$_{40}$As$_{40}$ alloy quenched from the melt.
Gray regions are remaining FeAs phase.
}
\label{f1}
\end{figure}

To check if the BaFe$_2$As$_2$ phase shows congruent melting or not, we have investigated microstructure of the BaFe$_2$As$_2$ alloy quenched from the melt.
Figure~\ref{f1} shows the resulting micrograph.
Apparently, the quenched alloy consists of the ternary BaFe$_2$As$_2$ phase with a small amount of the remaining FeAs phase.
The existence of the FeAs phase suggests that the initial melt composition may not be exactly BaFe$_2$As$_2$, most likely due to evaporation loss of the Ba element.
There is no trace of a peritectic reaction, {\it i.e.}, no trace of primary precipitates with other compositions, and the BaFe$_2$As$_2$ phase is definitely dominant.
Hence, we can conclude that most likely the BaFe$_2$As$_2$ directly forms from liquid of the identical composition, or in other words, it is a congruently melting phase.
The temperature could not be reliably measured in our induction furnace, however, we confirmed that the BaFe$_2$As$_2$ does not melt below 1443~K using standard electric furnaces.
Therefore, the melting temperature of the BaFe$_2$As$_2$ phase should be above 1443~K.

\section{Primary solidification field}
\begin{figure}
\includegraphics[scale=0.28,angle=-90]{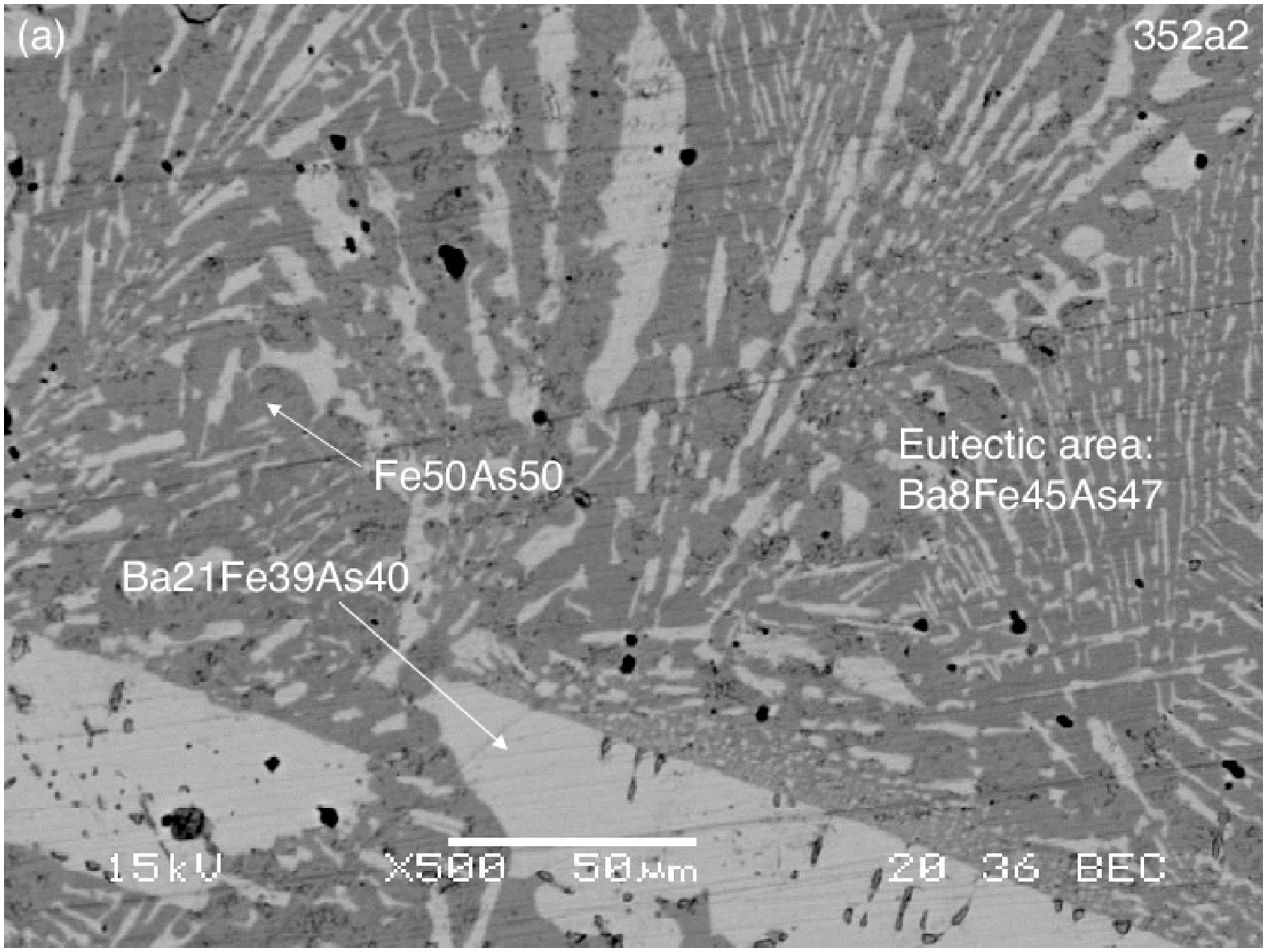}
\includegraphics[scale=0.28,angle=-90]{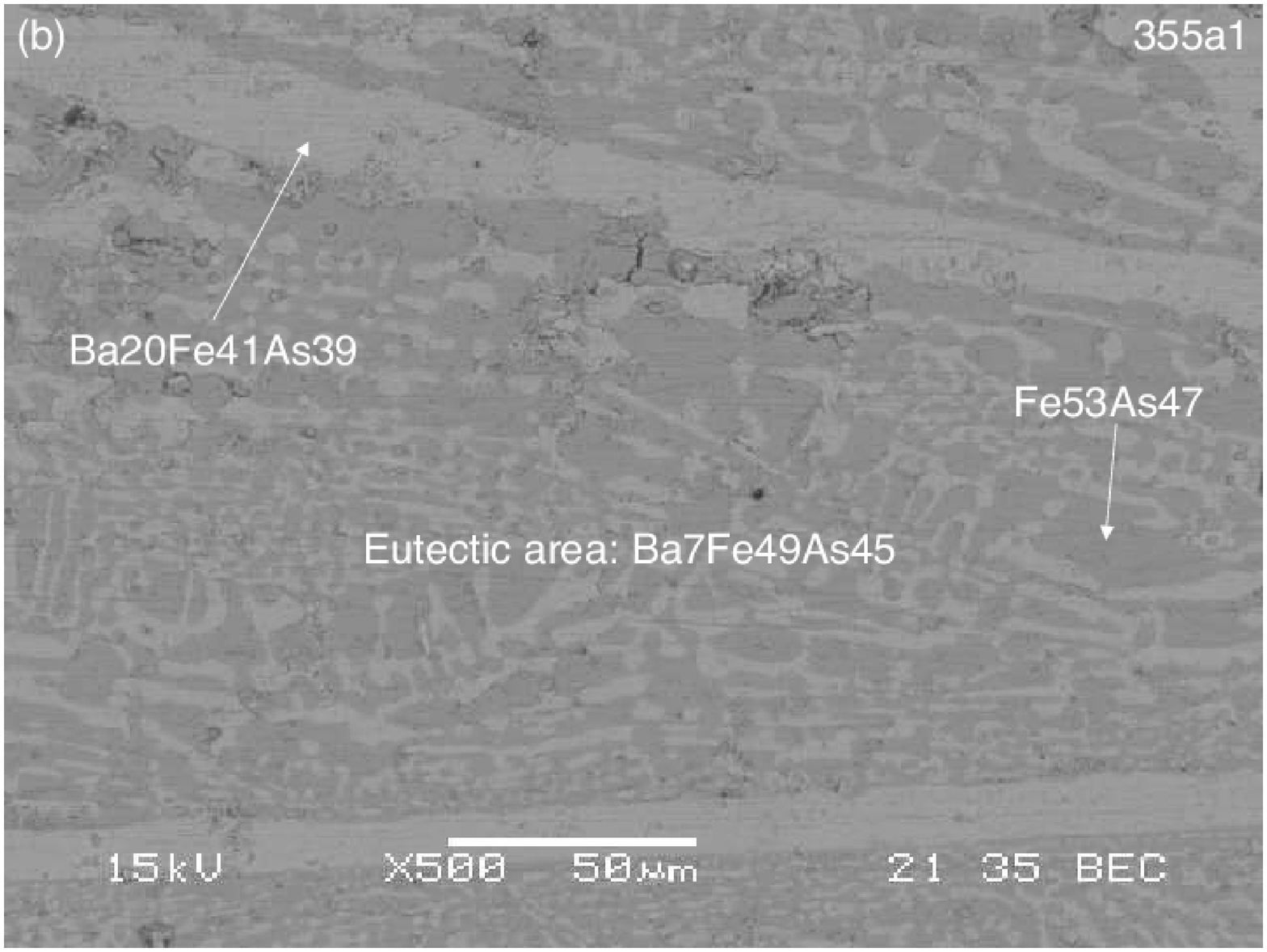}
\includegraphics[scale=0.28,angle=-90]{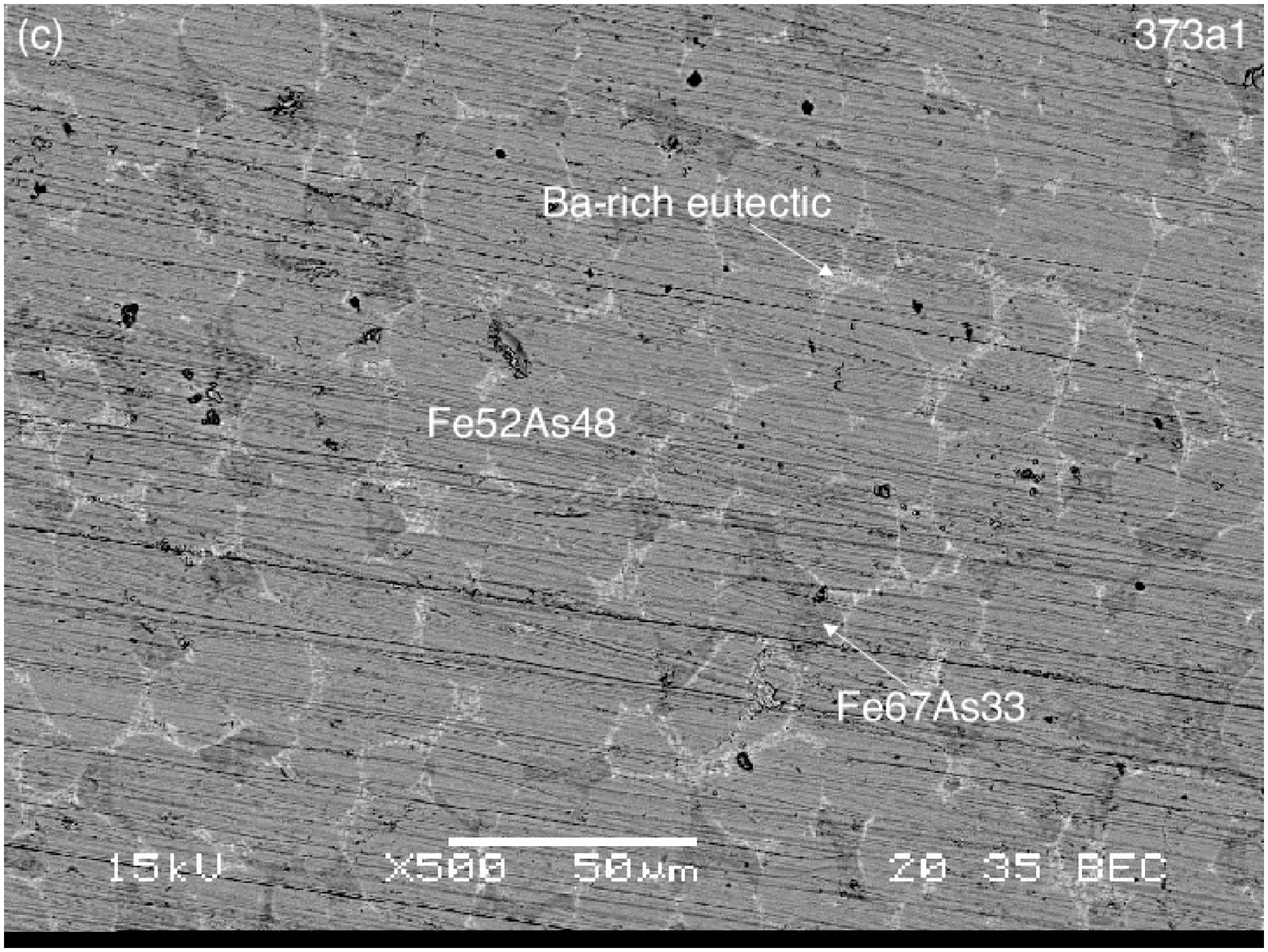}
\includegraphics[scale=0.28,angle=-90]{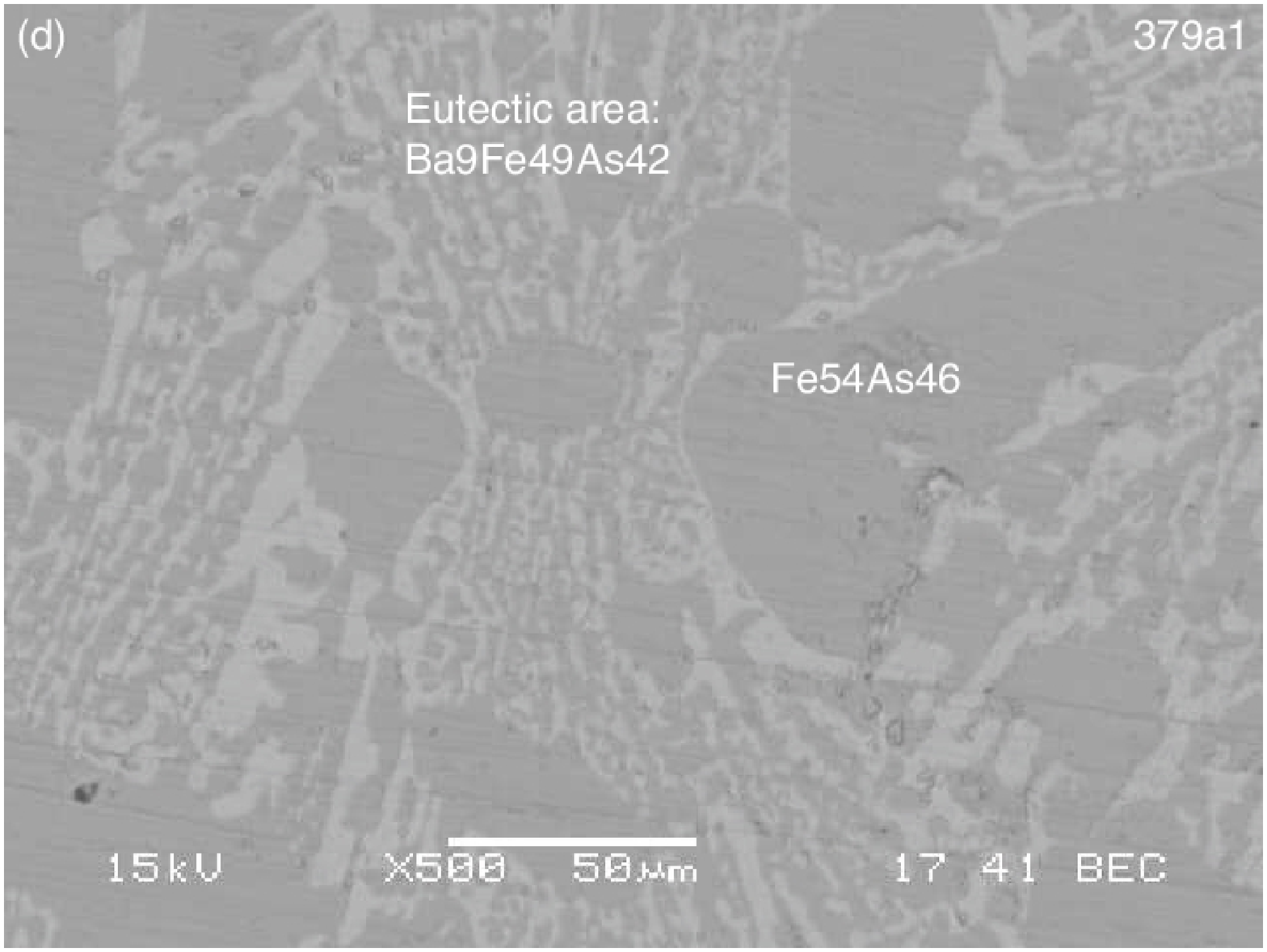}
\caption{Backscattered electron micrographs of the (a) Ba$_{15}$Fe$_{42.5}$As$_{42.5}$, (b) Ba$_{10}$Fe$_{45}$As$_{45}$, and (c) Ba$_{5}$Fe$_{47.5}$As$_{47.5}$ alloys annealed at 1423~K.
(d) Backscattered electron micrograph of the Ba$_{7.5}$Fe$_{46.25}$As$_{46.25}$ alloy annealed at 1323~K.
The brighter regions in the micrograph (d) correspond to the BaFe$_2$As$_2$ phase.
}
\label{f2}
\end{figure}

\begin{figure}
\includegraphics[scale=0.3,angle=-90]{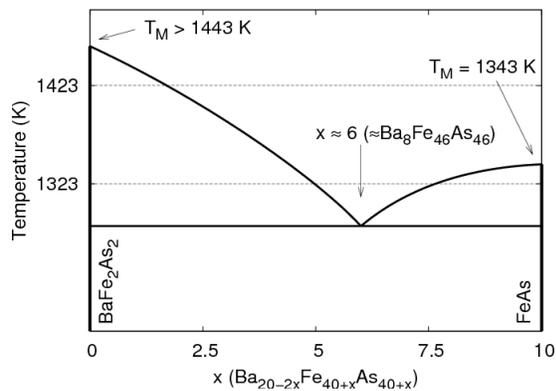}
\caption{Proposed pseudo-binary phase diagram along the Ba$_{20-2x}$Fe$_{40+x}$As$_{40+x}$ line.
 It should be noted that the melting temperature of the BaFe$_2$As$_2$ phase could not be determined in the present study.
The melting temperature of the FeAs phase is after Refs.~\protect\citeonline{selte72}.
}
\label{f3}
\end{figure}

\begin{figure}
\includegraphics[scale=0.28,angle=-90]{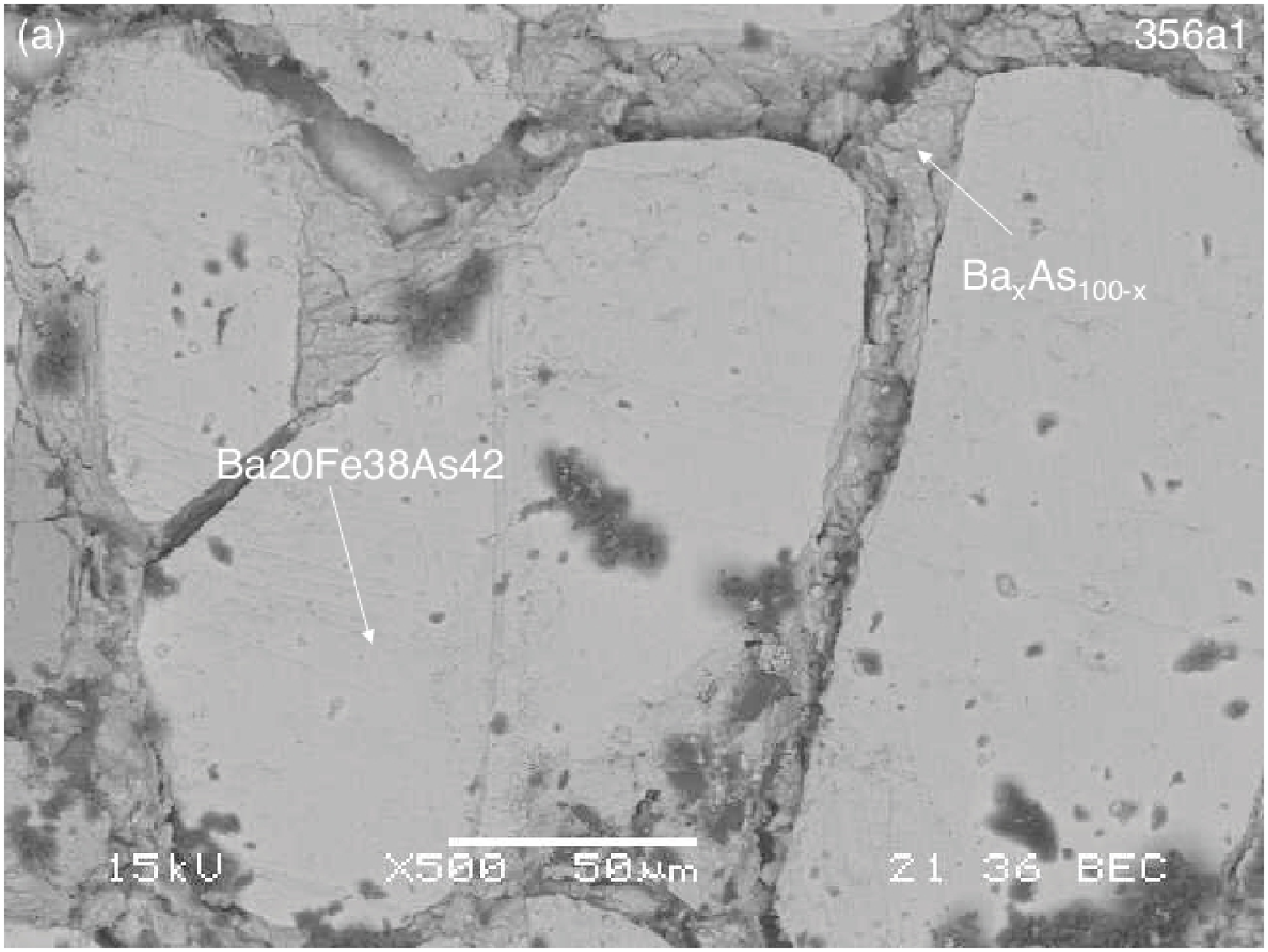}
\caption{Backscattered electron micrograph of the Ba$_{22.22}$Fe$_{33.33}$As$_{33.33}$ alloy annealed at 1423~K.
The composition of the liquid phase (Ba$_x$As$_{100-x}$) could not be determined reliably due to the roughness of the surface; the Ba-As phase was very unstable in the air, so that the phase was powderized when the polished alloy was transported to the SEM chamber.
}
\label{f4}
\end{figure}

Although we found that BaFe$_2$As$_2$ is most likely a congruently melting phase, it is favorable to achieve lower solidification temperature, as the conventional sealing using quartz tubes cannot survive at such high temperatures for a long time.
Therefore, we tried to find a flux which lowers the solidification temperature of the ternary phase. 
One way to reduce the solidification temperature is to use the self-flux method.
To optimize the self-flux crystal growth, we need to determine the primary solidification field of the BaFe$_2$As$_2$ phase in the Ba-Fe-As ternary phase diagram.
Therefore, we have investigated microstructure of alloys with variable compositions along the Ba$_{20-2x}$Fe$_{40+x}$As$_{40+x}$ and Ba$_{20+x}$Fe$_{40-3x}$As$_{40+2x}$ lines annealed at the two temperatures 1423 and 1323~K.

For the former Ba$_{20-2x}$Fe$_{40+x}$As$_{40+x}$ line, we have selected four compositions Ba$_{15}$Fe$_{42.5}$As$_{42.5}$ ($x = 2.5$),  Ba$_{10}$Fe$_{45}$As$_{45}$ ($x = 5$), Ba$_{7.5}$Fe$_{46.25}$As$_{46.25}$ ($x = 6.25$), and Ba$_{5}$Fe$_{47.5}$As$_{47.5}$ ($x = 7.5$).
Figure~\ref{f2}(a-c) shows the microstructures of the three alloys ($x = 2.5, 5$, and 7.5) annealed at 1423~K.
It should be noted that from shapes of the resulting ingots, those alloys were most likely melt at 1423~K.
For the $x = 2.5$ and 5 alloys, the BaFe$_2$As$_2$ phase appears in two different forms; one appears as large sized grains (typically larger than 200~$\mu$m in one direction), whereas the other forms very fine eutectic structure with the FeAs phase.
This indicates that there is a eutectic reaction between the BaFe$_2$As$_2$ and FeAs phases; the eutectic composition may be determined from the area scan in the EDX analysis as approximately Ba$_{8}$Fe$_{45}$As$_{47}$.
On the other hand, as shown in Fig.~\ref{f2}(c), the FeAs phase becomes dominant in the $x = 7.5$ alloy.
From the microstructure, we conclude that FeAs is the primary precipitate and the Ba-rich eutectic appears at the boundaries of the FeAs grains.
Therefore, $x = 7.5$ (Ba$_5$Fe$_{47.5}$As$_{47.5}$) is at the opposite side of the eutectic composition.
We have further prepared the Ba$_{7.5}$Fe$_{46.25}$As$_{46.25}$ ($x = 6.25$) alloy annealed at 1323~K.
Melting of this $x=6.25$ alloy at 1323~K was confirmed by the shape of the resulting ingot.
Microstructure of the resulting alloy is shown in Fig.~\ref{f2}(d); eutectic microstructure with FeAs and BaFe$_2$As$_2$ phases is clearly observed in the micrograph.
Therefore, we may conclude that eutectic temperature is below 1323~K.
These results provide the schematic pseudo-binary phase diagram as shown in Fig.~\ref{f3}.

Along the Ba$_{20+x}$Fe$_{40-3x}$As$_{40+2x}$ line, we checked the Ba$_{22.22}$Fe$_{33.33}$As$_{44.44}$ ($x = 2.22$) alloy.
The resulting micrograph of the alloy annealed at 1423~K is shown in Fig.~\ref{f4}.
It is clearly seen that the Ba-As liquid co-exists with the BaFe$_{2}$As$_{2}$ phase.
From the microstructure, the BaFe$_2$As$_2$ phase seems to be a primary precipitate in the Ba-As flux, and thus this composition may also be a candidate for the single-crystal growth.
However, as we found that the Ba-As phase is very unstable in the air, and is highly reactive with the Al$_2$O$_3$ crucible, we did not use this flux.

\section{Single-crystal growth}

\begin{figure}
\includegraphics[scale=0.28,angle=-90]{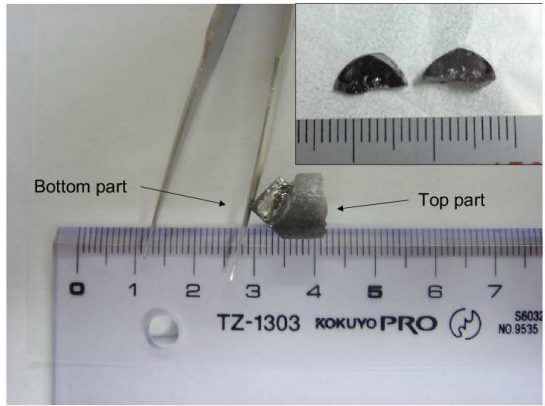}
\caption{A photograph of the resulting Bridgman-grown ingot.
The bottom part, where the growth of the single grains was confirmed, was broken into pieces on removal from the crucible.
Inset: the two pieces of the single grains obtained from the bottom part of the ingot.
Indeed, the two pieces came from the same grain, but were torn in parts when they were removed.
The thickness of each piece is about 2~mm.
}
\label{f5}
\end{figure}

\begin{figure}
\includegraphics[scale=0.28,angle=-90]{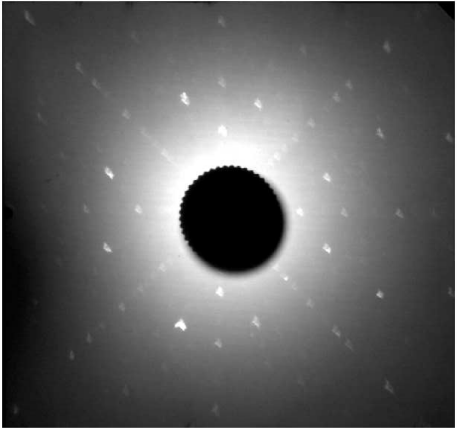}
\caption{X-ray Laue backscattering diffractogram of the obtained single crystal.
X-ray incidents along the normal to the cleaved surface.
Clear four-fold symmetry is seen in the diffractogram, indicating that the surface-normal corresponds to the $c$-axis.
}
\label{f6}
\end{figure}

\begin{figure}
\includegraphics[scale=0.3,angle=-90]{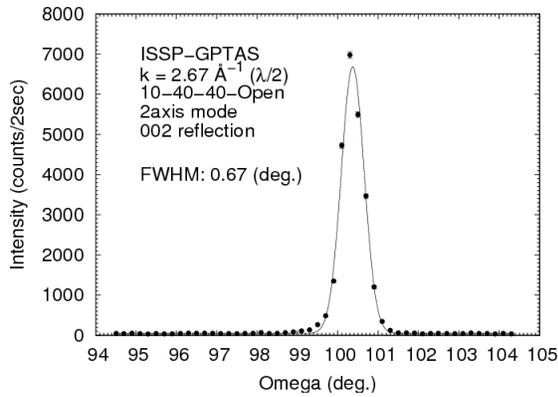}
\caption{$\omega$-scan around the 002 nuclear Bragg reflection of the obtained BaFe$_2$As$_2$ single crystal.
Solid line denotes a fitting to a gaussian, yielding the full width at half maximum (FWHM) of 0.67(1) degree.
}
\label{f7}
\end{figure}

\begin{figure}
\includegraphics[scale=0.3,angle=-90]{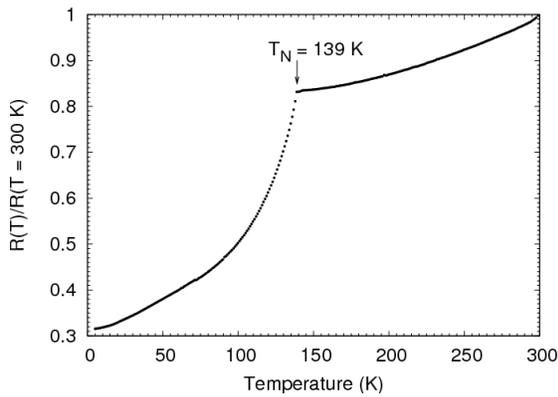}
\caption{Temperature dependence of the electrical resisitivity in the $ab$ plane.
An excitation current of 1~mA was used.
The resistivity shown here is normalized by the room-temperature value.
}
\label{f8}
\end{figure}

The vertical Bridgman technique was used to grow large-sized single crystals of the BaFe$_2$As$_2$ phase.
Based on the phase diagram study, we selected Ba$_{15}$Fe$_{42.5}$As$_{42.5}$ as the initial composition of the liquid.
After several trials with different growth speeds and initial temperatures, we found that reasonably large single grains (weighted up to 0.4~g) may be obtained with the speed of 0.5~mm/h and the initial temperature of 1443~K.

A photograph of the resulting ingot is shown in Fig.~\ref{f5}; the ingot was partly broken when it was removed from the Al$_2$O$_3$ crucible.
Single crystalline parts were found at the lower part of the ingot, as expected for the Bridgman growth.
However, we note that the bottom part was not entirely the single crystal; there are a few mis-oriented grains.
This suggests that the present Bridgman growth is not optimal; much lower growth speed and/or steeper temperature gradient is necessary to obtain an entirely single-crystalline ingot.

From the bottom part, two single crystalline pieces were removed as shown in the inset of Fig.~\ref{f5}.
The two grains were originally a part of one large crystal; it was easily be fractured on removal.
The size of the resulting grains is approximately 10~mm x 4~mm x 2~mm, which is an order of magnitude larger (mainly in thickness) compared to the earlier reports.

X-ray Laue backscattering diffractogram from the cleaved surface is shown in Fig.~\ref{f6}.
The four-fold symmetry can be readily seen, indicating that the cleaved surface indeed corresponds to the $hk0$ plane.
Figure~\ref{f7} shows the result of the $\omega$-scan of the crystal around the $002$ reflection, using neutron diffraction.
A very nice single peak feature was observed.
Since neutron penetrates deeply into the sample, this clearly shows that there are no accompanying mis-oriented grains in the obtained crystal.

Figure~8 shows the result of the electrical resistivity measurement.
A clear anomaly was observed at $T_{\rm N} = 139$~K, where the structure and the spin-density-wave transitions were known to take place~\cite{rotter08a}.
It should be noted that the resistivity ratio $R(T = 5~{\rm K})/R(T = 300~{\rm K}) \simeq 3$ is relatively small, however, is the same order of magnitude with that in the SrFe$_2$As$_2$ single crystal~\cite{chen08}.

\section{Summary}
We have performed the microstructure investigation on the ternary Ba-Fe-As system around the BaFe$_2$As$_2$ phase.
It was shown that the BaFe$_2$As$_2$ most likely melts congruently.
We also have determined pseudo-binary phase diagram along the Ba$_{20-2x}$Fe$_{40+x}$As$_{40+x}$ line, whereas supplemental information on the Ba$_{20+x}$Fe$_{40-3x}$As$_{40+2x}$ line was provided.
Based on the obtained pseudo-binary phase diagram, we have performed the vertical Bridgman growth of the BaFe$_2$As$_2$ phase, and obtained large crystals up to 0.4~g.
The quality of the resulting crystals was confirmed by the X-ray Laue and neutron diffraction.
Further optimization of the Bridgman growth parameters and trials to grow doped superconducting crystals are in progress.

\section*{Acknowledgment}
The present authors thank K. Ohgushi, N. Katayama, S. Ohhashi, and A. P. Tsai for valuable discussions.
This work has been partly performed using the facilities of MDCL, ISSP, University of Tokyo.


\end{document}